# Gravitational Hertz experiment with electromagnetic radiation in a strong magnetic field

N I Kolosnitsyn[1] and V N Rudenko[2,3]

[1]Smidt Earth Physics Institute RAS, *kolosnitsynn@mail.ru* , Moscow 119810, Russia,
[2]Institute of Nuclear Researches RAS, <*valentin.rudenko@gmail.com*> , Russia
[3]Sternberg Astronomical Institute MSU, Universitetskii pr. 13, Moscow 119234, Russia.

**Abstract.** Brief review of principal ideas in respect of the high frequency gravitational radiation generated and detected in the laboratory condition is presented. Interaction of electro-magnetic and gravitational waves into a strong magnetic field is considered as a more promising variant of the laboratory GW-Hertz experiment. The formulae of the direct and inverse Gertsenshtein-Zeldovich effect are derived. Numerical estimates are given and a discussion of a possibility of observation of these effects in a lab is carried out.

**Introduction**

Gravitational wave predicted by general relativity (see classical Einstein's paper [1]) remains to be the last type of radiation without a direct experimental confirmation. Although the indirect evidence of gravitational wave emission exists in astrophysics (a famous reduction of the orbital period of the Tailor's pulsar [2]) the aspiration to discovery and possession of this new form of matter turns up in the main stream of modern experimental physics. After the first attempts of J.Weber on detecting gravitational radiation of extraterrestrial origin [3] this branch of physical experiment has got an enormous development. Unique gravitational detectors - large solid state cryogenic resonance bars and long based optical interferometers at suspended mirrors have been constructed and passed through series of observation runs in expectation of a gravitational wave astrophysical signal. Recent reports of physical and technical aspects of these experiments one can find for example in the review papers [4] (physical results) and [5] (technical issue). However the main sequence of these experiments can be summarized in the short phrase: "up to now no any signals of gravitational wave nature were detected". At present gravitational interferometer detectors (LIGO, VIRGO) go through the stage of modernization with a goal of improving sensitivity at one order of value meanwhile the cryogenic bars (Nautilus, Auriga) continue to be in a duty cycle. This racing for the permanent increase of gravitational detector sensitivity stimulated by astrophysical forecast of the rate of random signals already is resulted in the extraordinary technical instrumentation which allows to perform measurements at the level of quantum limit of accuracy [6]. Here we have to remark that theoretical base of possibility to perform such "quantum measurements" were developed also in the papers of V.Manko with collaborators [7-9]. In these papers a rigorous definition of "quantum non demolition variable" (i.e. the variable which is free from quantum limitation) was introduced and a parallel with classical theory of optimal filtration was demonstrated. This understanding becomes especially important now for the successful development of the new (third) generation of the gravitational wave detectors [10]. At the same time a delay with "discovery of astrophysical gravitational waves" makes it reasonable to come back to analysis of the old idea of the "laboratory gravitational Hertz experiment".

It is interesting that the first consideration of the Hertz experiment with gravitational radiation in laboratory was carried out again by J.Weber in his pioneer paper [11]. He found that a generated power might reach $10^{-13}$ erg/sec but remained to be much less the receiver sensitivity. Afterwards several different mechanical and electromagnetic schemes were analyzed [12-15]. Brief summary of Russian researcher's contribution see in the papers [16,17]. Main conclusion was that in laboratory conditions of "slow motion" and "weak field" it would be extremely difficult to construct any effective gravitational generator and Hertz experiment seemed more as Gedanken figure then practical one. On the other hand an understanding that a gravitational radiation noise background at the typical radio VHF range $10^7 - 10^8$ Hz also must be negligibly small allows to continue a speculation on possibility of a "gravitational transmission of information" even with weak power transmitters

Modern programs for astrophysical GW searching using both ground-based and space apparatus deal with a frequency range does not exceed a few kilocycles per second. The upper frequency limit for relativistic sources is determined by the inverse fly-by time for a strong field region i.e. $\nu \leq (c/r_g) \approx 30 \cdot (3M_O/M) \cdot kHz$. The lower the frequency, the more intensive GW bursts one could expect according to the modern astrophysical forecast; in other words, massive relativistic stars are 'too heavy' and 'too inertial' for the production of powerful high frequency radiation. Nevertheless the high frequency gravitational radiation sources in astrophysics also were studied but this list is restricted to four different classes of objects: thermal gravitational radiation of stars [18] mutual conversion of electromagnetic and gravitational waves in a magnetized interstellar plasma [19], a relict cosmological gravitational wave background [20, 21] and gravitational radiation from very low mass primordial black holes [22]. Below we do not consider modified gravity models with large extra dimensions which predict profusive emission of light Kaluza-Klein gravitons (see for example the principal paper [23] ). However most part of these models is associated with violation of Equivalence Principle.

A brief review of high frequency sources in the frame of GR one can find in the paper [22]. Here we only point out some details of the relict gravitational wave background which will be used further in our article.

All theories give a non-thermal energy density spectrum for the GW relic background with a growth at low frequencies and a fall in the high-frequency region [20, 21]. It reflects a specific parametric mechanism of the GW amplification for gravitons with wavelengths of the order of the scale factor getting a more effective pump from the gravity field of the expanding world.

The spectrum at high frequencies has a cutoff approximately in the region $\nu_c \geq 10^{11} Hz$; it corresponds to the temperature, 0.9 K, that the GW relic background would have in the case of the adiabatic evolution [21]. Beyond $\nu_c$ the spectrum falls down very quickly: so at the frequency $10^{11} Hz$ the GW standard estimation is $h_\nu \approx 10^{-32} \cdot Hz^{-1/2}$ for the metric perturbation and $F \approx 10^{-5} - 10^{-6} \cdot erg.cm^{-2}s^{-1}$ for the GW flux. These data are very small in comparison with the sensitivity typical of modern gravitational detectors $h_\nu \approx 10^{-23} \cdot Hz^{-1/2}$ and with it at the much low frequency range $10^2 - 10^3 \cdot Hz$.

There are also estimates of stochastic high frequency GW background derived from the theory of primordial (big bang) nuclearsynthesis, so called "BBN constrain". The logic here is that a large GW energy density at the time of BBN would alter the abundances of the light nuclei produced in the process [24]. At low frequencies of LIGO and VIRGO detectors $(10^2 - 10^3) \cdot Hz$ the BBN constrain corresponds to spectral density of GW metric perturbation $h_\nu \approx 10^{-24} \cdot Hz^{-1/2}$. The BBN limit here was recently slightly beat in the LIGO S5 scientific run [25]. But at the more high frequency region $(10^{12} - 10^{15}) \cdot Hz$ BBN theory forecasts the much smaller limit $h_\nu \approx h_\nu \approx 10^{-34} - 10^{-37} \cdot Hz^{-1/2}$. Meanwhile the cosmic string model [26], and inflationary model [27] give estimate up to $h_\nu \approx 10^{-30} Hz^{-1/2}$. This the more optimistic forecast proposed by known theories we will have in mind during our consideration below in this paper.

Meanwhile for a laboratory generated gravitational radiation directed to a proper gravitational receiver for detection the high frequency range is just favorable one. The more promising frequency range for a gravitational Hertz test in a lab occupies the very high frequency interval $10^{10} - 10^{15} Hz$ [16, 17]. One stimulating idea here is based on the understanding that in a lab one could organize a coherent radiation from a large number of elementary quadruples (atoms or molecules of properly selected media, vortices in a superconductor, etc). The radiated power is expected to be relatively large because the very small factor of the quadruple formula G/c5 can be compensated by other factors: the enormous number of participating quadrupoles (up to $10^{22} - 10^{24} cm^{-3}$), sixth power of the oscillating frequency $\omega^6$, and a very sharp beaming of the output signal A composition of such type of multi cell setup of course is associated with a special selection of controversial parameters in respect of the maximum radiated power. Discussion and example of such optimization one can find in the paper [15-17].

It worth to note also that an elementary quadruple as a GW detector at very high frequencies should be more effective because its size can be matched to the gravitational wavelength (at low frequencies, bar detectors and even interferometers have the loose factor (mismatch) $(l/\lambda_g) \leq 1$).

In last years a new attention was attracted to the problem of the "laboratory Hertz experiment" with a high frequency gravitational radiation. (In particular the two special international conferences have been carried out in US on the subject: 1st HFGW conf., MIRT corp., McLean, May 6-9, 2003 and 2nd HFGW Int.WS, Institute for Advanced Studies at Austin (IASA), Texas, September 17-20, 2007, inspired by the main protagonist of the problem prof. R.M.L.Baker Jr). Partly it might be explained by a recent technological progress in nano structures manufacturing which promises a creation of very effective multi cell coherent GW-radiators in a form of relatively compact solid chips controlled by PC.

A practical interest is associated with the "old dream" to realize a new "low noise and extremely far distance" communication line which might be effective at space scale using the gravitational waves as most penetrative type of radiation in the nature.

In the microwave and optical region the magneto-optical principle of HFGW generation and detection was last years in the center of attention of research groups involved in the problem. The idea of a mutual EM-GW conversion in a magnetic field was initially proposed and estimated in the pioneer papers of Gertsenshtein [28] and Zeldovich [29,30] in application to some astrophysical phenomena. Recently this principle was explored in respect of the detection of high frequency tail of relic gravitational wave background and some construction of correspondent magneto-optical receiver was proposed [31]. In this paper we follow to the goal of independent estimation and checking a possibility of using the "direct and inverse Gertsenstein effect" for a laboratory construction of HFGW generator and detector. In distinguish of the paper [31] we use the well known optical instrument - FP cavity for amplification of the EM-GW and GW-EM conversion effects. Gravitational wave amplitude produced by a generator based at the direct Gertsenstein effect is calculated. Then this wave is sent at the "inverse Gertsenstein-Zeldovich receiver" and the registration problem is estimated at the black body radiation background. Finally we consider a principal possibility of application FP cavity to enhance the conversion effect and briefly discuss one of such type detector [31] proposed for a relic GW background

## 1. Basic equations

One can start from the Einstein equations in the following form

$$R_{ik} = \frac{8\pi G}{c^4}\left(T_{ik} - \frac{1}{2}g_{ik}T\right). \tag{1}$$

Just this form is convenient for an analysis of gravitational wave generation by a variable electromagnetic field so as the electromagnetic energy-momentum tensor of the field has a trace equal to zero ($T = 0$) and the equations (1) are reduced to

$$R_{ik} = \frac{8\pi G}{c^4}T_{ik}. \tag{2}$$

In particular below we will be interested with spatial components of the gravitational wave. Then the energy-momentum tensor in (2) can be replaced by the Maxwell stress tensor:

$$T_{ik} \to T_{\alpha\beta} = -\sigma_{\alpha\beta}, \quad \alpha, \beta = 1, 2, 3.$$
$$\sigma_{\alpha\beta} = \frac{1}{4\pi}\left\{E_\alpha E_\beta + H_\alpha H_\beta - \frac{1}{2}\delta_{\alpha\beta}\left(\boldsymbol{E}^2 + \boldsymbol{H}^2\right)\right\}. \tag{3}$$

The next simplification consists in transition to the weak gravitational field approximation

$$g_{ik} = \eta_{ik} + h_{ik}, \quad |h_{ik}| \ll 1$$

with the Lorentz gauge (or TT gauge [32])

$$\bar{h}^{ik}{}_{,k} = 0, \quad \bar{h}_{ik} = h_{ik} - \tfrac{1}{2}\eta_{ik}h,$$

Under this the Ricci tensor takes the form:

$$R_{ik} = \frac{1}{2} \Box h_{ik}.$$

Finally instead of (1) one comes to the equation

$$\Box h_{\alpha\beta} = -\frac{16\pi G}{c^4}\sigma_{\alpha\beta}. \qquad (4)$$

## 2. The Gertsenshtein – Zeldovich problem

We will use the equation (4) to get solutions for so called "direct Gertsenshtein – Zeldovich (GZ) problem", i.e. - a generation of gravitational wave by the electromagnetic one in the presence of constant magnetic field, and the inverse GZ problem, - a birth of electromagnetic field under action of metric perturbation (equivalent of GW action) in the strong magnetic field.

Let's start from a simple configuration of electromagnetic fields. The constant magnetic field $\boldsymbol{H} = \boldsymbol{e}_2 H_0$ along the y-axis covers the x-region: $0 \leq x \leq L$. Besides the electromagnetic wave with linear polarization along z-axis travels along the positive direction x-axis. Then a total electromagnetic field has the following two components

$$\tilde{\boldsymbol{E}} = \boldsymbol{e}_3 \tilde{E}_0 \cos(\omega t - kx), \qquad \tilde{\boldsymbol{H}} = \boldsymbol{e}_2(H_0 - \tilde{H}_0 \cos(\omega t - kx)). \qquad (5)$$

We prefer to use here the "Gaussian system of units" where the four main vector variables $E, D, B, H$ have the same dimensions (see Sommerfeld [33]). In particular the electric and magnetic components of the plane electromagnetic wave in this system became equal: $\tilde{E}_0(CGSE) = \tilde{H}_0(CGSM)$. The matrix of Maxwell stress tensor can be right down easily as

$$\sigma_{11} = -\frac{1}{8\pi}\left\{H_0^2 - 2H_0\tilde{H}_0\cos(\omega t - kx) + 2\tilde{H}_0^2\cos(2\omega t - 2kx)\right\}, \qquad (6)$$

$$\sigma_{12} = \sigma_{21} = 0,$$
$$\sigma_{13} = \sigma_{31} = 0,$$
$$\sigma_{23} = \sigma_{32} = 0,$$

$$\sigma_{22} = -\sigma_{33} = \frac{1}{8\pi}\left\{H_0^2 - 2H_0\tilde{H}_0\cos(\omega t - kx)\right\}, \qquad (7)$$

It is clear from (4)-(7) that the given combination of the electromagnetic fields has to generate a plane transversal gravitational wave with the polarization known as $h_+$. It is composed by two the diagonal components $h_{22} = -h_{33}(= h_+)$.

We have to remark here a special feature of this type conversion, namely, the appearance of a longitudinal component of GW $h_{11}$ which was produced by the diagonal component of the stress tensor $\sigma_{11}$. The existence of such polarization, besides the principle ones $h_+, h_{[x}$, is admitted in general relativity,( see [34]) but below we will be concentrated only at the transverse field components.

## 3. Calculation of the direct GZ-effect

The source of the gravitational radiation in the equation (4) is the Maxwell stress tensor, containing an EM harmonic wave and the constant magnetic field in the region $0 \leq x \leq L$. Let us seek a solution of (4) choosing for example the component of $h_{22}$. For linearly polarized EM the correspondent value of $\sigma_{22}$ (= - $\sigma_{33}$) is defined by (7) without the first constant term. Then the formula (4) is reduced to the equation

$$\left(\frac{\partial^2}{\partial x^2} - \frac{\partial^2}{c^2 \partial t^2}\right) h_{22} = \frac{4G}{c^4} H_0 \tilde{H}_0 \cos(\omega t - kx). \qquad (8)$$

Its simple solution can be found in the form

$$h_{22} = Ax \cos(\omega t - kx + \varphi).$$

A substituting these expressions in (8) results in the equation for parameters A, φ:

$$2kA \sin(\omega t - kx + \varphi) = \frac{4G}{c^4} H_0 \tilde{H}_0 \cos(\omega t - kx).$$

It follows from here

$$2kA = \frac{4G}{c^4} H_0 \tilde{H}_0, \quad \varphi = \frac{\pi}{2}.$$

Finally one comes to the metric component of generated GW as

$$h_{22} = -\frac{2G}{c^4 k} H_0 \tilde{H}_0 x \sin(\omega t - kx). \qquad (9)$$

The amplitude (9) linearly grows along the way. But the effective travel distance is restricted by the size of magnetic field region (interaction zone). So the maximum generated GW amplitudes are

$$\max h_{\alpha\beta} = \frac{2G}{c^4 k} H_0 \tilde{H}_0 L \qquad (10)$$

In principle a productivity of EM-GM conversion can be enhanced with a Fabry-Perot cavity placed in the region overlapped by magnetic field. At the cavity resonance $\mu = 2kL = 2\pi n$ the amplitude of resonating EM waves are described by the following expression [35]

$$\tilde{\mathbf{H}}^{\pm} = \mathbf{e}_2 \tilde{H}_0 \frac{1}{1 - r_1 r_2} \cos(\omega t \pm kx), \qquad (11)$$

where $r_1$ and $r_2$ mirror's reflection coefficients; signs (±) present two parts of standing wave inside the cavity. So inside the Fabry-Perot cavity the EM amplitude is increased in factor of $1/(1 - r_1 r_2)$. Putting (11) in (9) one can estimate the maximal amplitude of the gravitational wave generated in the Fabry-Perot cavity and traveled for example in the positive x direction

$$\max h_{\alpha\beta} = \frac{G}{c^4 k} H_0 \tilde{H}_0 \frac{2L}{1 - r_1 r_2} = \frac{G H_0 \tilde{H}_0 L \lambda_{em}}{c^4 \pi^2} F. \qquad (12)$$

Here $F = \pi/(1 - R)$ is the finesse, $R = r_1 r_2$ is a power reflectance of the mirrors ($r_1$ and $r_2 \sim 1$). Modern technologies allow a manufacturing of superconductive resonators with the quality up to the level of energy reflectance $R = r_1 r_2 \sim 1 - 10^{-8}$ and in future may be $1 - 10^{-10}$.

For numerical estimations let's take the involved parameters close to their technical limit: $H_0 = 10^5$ Gs (or 10 Tesla); $L = 100$ m $= 10^4$ cm, $\lambda = 1$ cm; $1 - R = 10^{-8}$; the EM amplitude $\tilde{H}_0$ can be derived from the pump power $P = \tilde{H}_0^2 cS / 4\pi$. Then for the cross section of EM beam $S = 1$ cm$^2$, and $P = 10$ kW (or $10^{11}$ erg/s) one obtains - $\tilde{H}_0 = 6{,}47$ Gs. Finally the estimation of generated GW amplitude results in $h_{\alpha\beta} \sim 1 \cdot 10^{-32}$. For advance superconductive microwave resonators with finesse F~ $10^{10}$ the final estima-tion could reach $h_{\alpha\beta} \sim 1 \cdot 10^{-30}$. (it's worth to note that for microwave resonator the parameter F roughly can be equivalently replaced by the resonator quality factor.)

### 4. Evaluation of the inverse GZ-effect.

Let's consider the "inverse GZ problem" i.e. a detection of the EM radiation generated by GW in the strong magnetic field. As above let the region $0 \leq x \leq L$ is filled by the constant magnetic field $\boldsymbol{H} = \boldsymbol{e}_2 H_0$ and the gravitational wave travels along the positive direction of x-axis. The goal is to estimate an amplitude of the electromagnetic wave resulted from the interaction of GW with magnetic field. One can proceed from the three-dimensional Maxwell equations written in a curved space (it reflects a presence gravitational field) [29, 36]:

$$\Box \boldsymbol{E} = \boldsymbol{F}_E, \quad \Box \boldsymbol{H} = \boldsymbol{F}_H, \tag{13}$$

where the right of (13) are specified as

$$\boldsymbol{F}_E = \frac{\partial}{c \partial t} \mathrm{rot}(\hat{h} \boldsymbol{H}_0),$$

$$\boldsymbol{F}_H = \frac{\partial^2}{c^2 \partial t^2}(\hat{h} \boldsymbol{H}_0) - \mathrm{grad\,div}(\hat{h} \boldsymbol{H}_0), \tag{14}$$

$$\hat{h} = \{h_{\alpha\beta}\}, \quad \alpha, \beta = 1, 2, 3. \quad \Box = \Delta - \frac{\partial^2}{c^2 \partial t^2}.$$

In the field configuration formulated above the equation for electric component reduced to

$$\Box \boldsymbol{E} = \frac{\partial}{c \partial t} \mathrm{rot}(\hat{h} \boldsymbol{H}_0). \tag{15}$$

Here

$$\mathrm{rot}(\hat{h} \boldsymbol{H}_0) = \mathrm{rot}\left(\begin{bmatrix} 0 & 0 & 0 \\ 0 & h_{22} & 0 \\ 0 & 0 & -h_{22} \end{bmatrix} \boldsymbol{e}_2\right) H_0 = H_0 \mathrm{rot}\boldsymbol{h},$$

$$\boldsymbol{h} = \begin{pmatrix} 0 \\ h_{22} \\ 0 \end{pmatrix} = \boldsymbol{e}_2 h_{22} = -\boldsymbol{e}_2 \frac{2G}{c^4 k} H_0 \tilde{H}_0 x \sin(\omega t - kx),$$

$$\mathrm{rot}(\hat{h} \boldsymbol{H}_0) = H_0 \mathrm{rot}\boldsymbol{h} = -\boldsymbol{e}_3 \frac{2G}{c^4 k} H_0^2 \tilde{H}_0 [\sin(\omega t - kx) - kx \cos(\omega t - kx)] \tag{16}$$

Substituting the expression (16) in (15) and assuming $\boldsymbol{E} = -\boldsymbol{e}_3 E$ we obtained the equation for the electric field $E$, :

$$\Box E = \frac{2G}{c^4} H_0^2 \tilde{H}_0 [\cos(\omega t - kx) + kx \sin(\omega t - kx)] \tag{17}$$

We interest the second term in the right side with amplitude $kx \gg 1$ the EM wave length is much less a size of a magnetic field region. Then the formula (17) is reduced to the equation

$$\left(\frac{\partial^2}{\partial x^2} - \frac{\partial^2}{c^2 \partial t^2}\right) E = \frac{2G}{c^4} H_0^2 \tilde{H}_0 kx \sin(\omega t - kx). \tag{18}$$

A solution of the equation we seek in the form

$$E = A\left[x\sin(\omega t - kx + \varphi) + kx^2 \cos(\omega t - kx + \varphi)\right] \tag{19}$$

Further we find the second derivatives of $E$ with respect to $x$ and respect to $t$:

$$E''_{xx} = A\left[3k^2 x\sin(\omega t - kx + \varphi) - k^3 x^2 \cos(\omega t - kx + \varphi)\right],$$
$$\ddot{E}_{tt} = A\left[-k^2 x\sin(\omega t - kx + \varphi) - k^3 x^2 \cos(\omega t - kx + \varphi)\right].$$

From here

$$\Box E = E''_{xx} - \ddot{E}_{tt} = 4Ak^2 x\sin(\omega t - kx + \varphi).$$

Comparing this quantity with the equation (18) we obtain

$$A = \frac{G}{2c^4 k} H_0^2 \tilde{H}_0, \qquad \varphi = 0. \tag{20}$$

As the result

$$E = \frac{G}{2c^4 k} H_0^2 \tilde{H}_0 \left[x\sin(\omega t - kx) + kx^2 \cos(\omega t - kx)\right] \approx$$
$$\approx \frac{G}{2c^4} H_0^2 \tilde{H}_0 L^2 \tag{21}$$

It is the maximum amplitude of the signal EM wave which it can reach at the output of magnetic field region. However in principle the induced EM radiation can be enhanced by the Fabry-Perot cavity placed along the x-axis normal to the magnetic field. Inside the resonator with high reflecting mirrors the electromagnetic field will be increased in the factor 1/(1-R). However the problem consists in a way of measurement of the electromagnetic field accumulated inside the cavity. One of ideas discussed in literature is a using a cavity with modulated quality factor [37]. A sharp falling down of the reflectivity of the output cavity mirror would produce a short pulse of accumulated EM energy which might be detected by a sensitive photo sensor. In particular the utilization for this purpose a fractal mirror with derived reflectivity was proposed in the paper [31, 38, 39] but technical details have to be addressed.

As for the expected EM quantum flux density induced by the coherent gravitational wave it can be estimated through the formulae (23)-(24) amplified by the quality factor of FP cavity. Then for EM quantum flux $N_S$ after integration of the output power over the FP cavity cross-section S, one can get

$$N_S = \frac{\lambda h^2 H_0^2}{2\hbar(1-R)^2}\left(\frac{L}{\lambda}\right)^2 S \tag{22}$$

For a numerical estimation one has to apply the parameters close to those were used above for the problem of GW generation:

$$h = h_+ = h_\times = 10^{-32}, \quad L = 100 \text{ m} = 10^4 \text{ cm}, \quad H_0 = 10^5 \text{ Gs (10 T)}, \quad \lambda = 1 \text{ cm},$$
$$1 - R = 10^{-8}, \quad S = 1 \text{ cm}^2.$$
.

Substitution these numbers in the formula (22) yields $N_S \approx 10^{-3} \text{ s}^{-1}$. It seems it is too small flux to be registered. For to improve situation one has to admit the FP resonator having the finesse (or quality factor) $F = \pi/(1-R) \approx 10^9$. Then generated GW amplitude will have the order (12) $h \approx 2 \cdot 10^{-31}$, and the signal photon flux resulted in $N_S \approx 14 \cdot s^{-1}$. Could this flux be registered ?

Preliminary answer to this question on principal level might be found through comparison of the signal with the noise produced by the black body radiation background. The number of background photons in the volume $V$ with frequency $\nu$ inside the bandwidth $\Delta\nu$ is presented by the classical formula [40]

$$N = \frac{V 8\pi}{c^3} \frac{\nu^2 \Delta\nu}{\exp\left(\frac{h\nu}{kT}\right) - 1} \tag{23}$$

Substitution $\nu = 3 \cdot 10^{10} Hz$, the bandwidth of the resonator $\Delta\nu \approx \nu/F \approx 30 \cdot Hz$, the resonator volume $V = 10^4 cm^3$ and the temperature $T = 300 \cdot K$ gives the number of black body photons $N \approx 0.05 \cdot$. For the cryogenic resonator with $T = 4K$ the expected number of photons is $N \approx 0.006$. One has to take into account that inside the resonator a total number of background photons is increased in F times (or in a quality factor), i.e. for the cryogenic case this number has to be $N_{tot} = NF \approx 6 \cdot 10^6$. Now one can estimate the measurement time required for the generated number photons would exceed the background

$$\tau_{mes} = \frac{N_{tot}}{N_S} \approx 4 \cdot 10^5 s \approx 4\, days \tag{24}$$

This estimation presents only so called "potential level of sensitivity", which ignores many different instrumental noises and technical problems. In the real experiment a long measurement time may be equivalently replace by a correspondent narrow band filtering.

### 5. Discussion

Above a simple combination (couple) of the gravitational wave generator and receiver operating through mechanism of the direct and inverse GZ-effect was analyzed and supplied by correspondent numerical estimations . It was shown that for the extreme but not too fantastic parameters of the setup it would be possible to satisfy (in principle) the condition of the Gravitational Hertz experiment i.e. a radiated GW power is turned out to be sufficient for registration by the "inverse receiver".

It was occurred at the value of GW amplitude on the order of $h \approx 10^{-30} - 10^{-32}$. This result appears due to application of FP cavity for an enhancing the amplitude of EM carrier in generator and EM signal wave in the receiver (in the paper [31] this possibility was not taken into account).

Meanwhile originally the direct Gertsenshtein effect was formulated for an infinite open space [23] as well as the inverse effect [24]. Thus a physical argumentation in favor of a possibility of FP cavity using has to be given. It might be presented as follows;

    a) direct effect.

In the free space zone fulfilled by the magnetic field a birth of gravitational wave occurs due to the time dependent quadruple term in the right part of the equation (8) produced by the travelling electromagnetic wave. Synchronism of the process is provided by the equality of frequencies and wave vectors of the both waves. Insertion of FP cavity mirrors with the base L changes the EM wave dynamics. The standing wave is created with amplitude increased in the finesse factor F equaled roughly to the number of EM wave round trips between mirrors. The standing wave formally can be considered as the sum of two travelling waves running in opposite directions along the x-axis. Each of these waves consists in F segments. Between the neighbor wave segments running for example to the right there is the time shift 2L/c and the phase shift 360 grad. So their sum reproduces continuous EM wave (a summing of different wave peaces (teams) with time shift L/c) travelled the distance F(L/c) through the space filled in by the magnetic field. The correspondent gravitational wave amplitude generated due to the Gertsenshtein effect will be proportional to ~ $EH_0$ (FL/λ). The equivalent coherent length of the laser radiation has to be larger the distance 2FL. These requirements will be fulfilled under the optical resonance condition L=n (λ/2). The gravitational wave rises in the form of two anti phase components leaving the source (FP cavity) in the opposite directions of x-axis.

    b) inverse effect

In the magnetic field irradiated by continuous gravitational wave a coherent EM radiation has to be born. Its amplitude must be increased with multi passes through the magnetic field zone. As above let's concentrate our attention on EM wave segments inside the FP cavity travelling along the direction of gravitational wave. Under the condition of FP resonance these wave segments will be coherently summarized resulting in enhancing the amplitude in F(L/λ) times It is just equivalent to the inverse Gertsenshtein- Zeldovich effect in a free space with the base FL filled by magnetic field. There is no any compensation with untiphase wave segments because the last ones have the time shift (L/c) in respect of the inphase segments. As well the total amount of EM waves produced in the same magnetic field zone can not have a mutual compensation so as the inphase and antiphase waves come out in opposite directions and do not interfere with each other.

In the physical picture presented above it was supposed the absence of any additional phase shifts of EM waves in magnetic field besides the shift at π grad under reflection from FP mirrors, i.e. there were no nonlinear effects type of optical birefringence, rotation of polarization plane etc.

A detailed elaboration of the technical scheme of the experiment was not a subject of anxiety in this paper. It was concentrated on the investigation of principal opportunity to perform such experiment with the present level of experimental technique. Above only main parts of the experimental scheme in one dimension structure were defined. In that number are i) GW generator as a microwave FP cavity in the strong magnetic field with external EM power pump; ii) EM receiver isolated from this pump also as FP cavity in the similar magnetic field; iii) the registrar of the output signal as a sensitive radiometer (photodetector).

Of course our positive conclusion about a feasibility of electromagnetic variant of the Hertz experiment takes place only at the level of "potential opportunity" (it was supposed that all generated GW radiation directly hits to the receiver, other losses also are absent etc.). To compose any realistic engineering scheme typical questions such as a supply power, coherent hindrances, budget of noises and achievable signal-to-noise ratio have to be addressed.

At the end we would like make some comments in respect of a high frequency gravitational wave detector with an attractive sensitivity proposed in the paper [31]. It explores not the pure inverse GZ effect but introduces some modification adding to the strong magnetic field also an "EM Gaussian beam" along the GW propagation axis. It is supposed that interaction of this beam with destined GZ photons at the area of some fractal mirror [38, 39] produces an amplification of the secondary (signal) GZ photons. It resulted in the detector efficiency proportional to the first power of gravitational wave amplitude and electromagnetic amplitude of the Gaussian beam. According to our estimations and earlier estimation for one special case considered by Zel'dovich [30], the detector sensitivity has to be proportional to the square of gravitational wave amplitude. Also the requirement of resonance wave synchronism in the process of conversion of gravitational waves into electromagnetic ones was not completely clear for the specific of the scheme [31] and also has to be addressed. But our main doubt is associated with the circumstance that this detector was proposed for a detection the relic gravitational wave background at its high frequency tail with stochastic standard estimation $<h> \approx 10^{-30} \cdot Hz^{-1/2}$

As we have seen this level corresponds to the very optimistic detection threshold even for the deterministic signal with well know frequency and phase, so a registration of the stochastic background looks a bit unbelievable. Moreover a detection of very weak electromagnetic perturbations at such threshold level unavoidably has to be associated with any Quantum Electromagnetic Detection devices [6] and measurement procedure type of Quantum Non demolition Measurements [41]

**Acknowledgement**

Authors thank colleagues from RAS professors I D Novikov and G S Bisnovatyi-Kogan for fruitful discussions and also dr A.V.Gusev for consultations on radio physical aspects of the problem
In our final remarks we would like express our sincere acknowledgement to professor
Vladimir Man'ko for his assistance (during many years of our scientific collaboration) in understanding of a crucial role of the quantum oscillator in the modern physics in general and in the theory of precise measurements in particular. Also we gratitude his wife Margaret Man'ko for her permanent help in the process of preparation of articles submitted to IOP magazines. We wish to both a good health and a long time creative scientific activity.
.

**References**

[1] Einstein A 1918 Sitzungsher. Preuss. Akad. Wiss.,**1** 154
[2] Taylor J H  1994  *Rev.Mod. Phys.* **66** 711
[3] Weber J 1968  *Phys. Rev. Lett*. **20** 1307
[4] Bizouard M A   and  Papa M A. Searching for gravitational waves with the LIGO and Virgo inter- fereometers. *arXiv:1304.4984v1*  17 Apr 2013
[5] Adhikari R X  2014  *Reviews of modern physics*. **86** 121
[6] Braginsky V B , Vorontsov Yu I  1974  *Sov.Phys.Uspekhi*  **114**  41
[7] Dodonov V V, Man'ko V I , Rudenko V N   1980 *Sov.Phys. JETP*  **81**  443
[8]  . Dodonov V V, Man'ko V I , Rudenko V N  1982  *Zhurn.Exp.Teor.Fiz Pisma* **36**  (iss 3)  53
[9]   Dodonov V V, Man'ko V I , Rudenko V N   1983 *Foundation of Physics*, **13**,  (iss 8)   607
[10] Acernese F *et al*   2008 *Class. Quantum Grav*.  **25**  184001
[11]  Weber J  1960  *Phys.Rev.***117**  306
[12]  Halpern L. and Laurent B  1964 *Nuovo Cimento* **33** 728
[13]  Kopvillem U Kh and Nagibarov V R   1969 *JETP Lett.*  **2**  (iss 12) 329
[14]   Grishchuk L P , Sazhin M V 1975  *Sov.Phys. JETP*   **41** 787
[15]   Braginsky V B and Rudenko V N  1978   *Phys. Rep.* **46** 165
[16] Rudenko V N  2003   Optimization of parameters of a couple generator-receiver
    for a gravitational Hertz experiment  *arXiv: gr-qc/0307105*
[17] Rudenko V N  2004  *Gravitation & Cosmology*  **10**, (iss 1-2)  41
[18]   Weinberg S  1972 Gravitation and Cosmology (New York: Wiley)
[19]  Galtsov D V  1975  *Sov. Phys. JETP*  **40** 211
[20]  Grishchuk  L P  1988  Usp. Fiz. Nauk  156  297
[21]  Allen B 1997  Relativistic Gravitation and Gravitational Radiation ed J A Marck and J P Lasota
    (Cambridge: Cambridge University Press) p 373
[22]   Bisnovatyi-Kogan G S and Rudenko V N  2004  Classical Quantum Gravity  **21**   3347
[23]   Arkani-Hamed N,  Dimopoulos S,  Dvali G.R.  1999  Phys.Rev.D  **59**  086004
[24] M. Maggiore M. 2000  Phys. Rept. **331** 283
[25] B. Abbott, et al., 2009  Nature 460 990
[26] Siemens X, Mandic V, Creighton J,  2007 Phys. Rev. Lett. 98 111101
[27] Starobonskii A.A. 1979 JETP Lett.  30, 682
[28]  Gertsenshtein M E  1962  *Sov. Phys. JETP Lett.* **14** (iss 1) 84
[29] Zeldovich Ya B 1973 *Sov. Phys. JETP* **65** 1311
[30]  Zeldovich Ya.B. and Novikov I D 1975 Structure and Evolution of the Universe. (Moscow: Nauka)
[31] Li F, Yang N, Fang Z,  Baker Jr R M L, Stephenson G V, Wen H  2009  *Phys. Rev. D*  **80** 064013
[32] Landau L. and Lifshitz E.M 1962 The Classical Theory of Fields (New York: Pergamon).
[33]   Sommerfeld A 1949  Elektrodynamik  Leipzig
[34]   Newman E, Penrose R 1962 *J. Math.Phys.*  **3**  566
[35]  Vinet J-Y  1986  *Physique*, **47**, 639-43
[36]  Kolosnitsyn N I  1996  *Gravitation & Cosmology*  **2** (iss 3)  262
[37]  Besotosnii V., Cheshev E., Gorbunkov M., Kostryukov P., Krivonos M., Tunkin V.,
    Jakovlev D. Applied Physics B: Lasers and Optics (2010) **101** 71
[38]  Wen W, Zhou L, Li J, Ge W, Chan C and Sheng P..2002  Phys. Rev. Lett**. 89**, 223901
[39]  Zhou L,  Wen W,  Chan C and Sheng P. (2003) Appl.Phys.Lett. **82** (7) 1012
[40] Landau L.D., Lifshitz E.M. 1980 The Statistical Physics part 1 (Elsevier Ltd Oxford)
[41]   Braginsky V B.  2007  PNAS  104  3677-80